\documentclass[prl,twocolumn]{revtex4}
\usepackage{amsfonts}
\usepackage{amsmath}
\usepackage{psfrag}
\usepackage{amssymb,amsmath,amsthm}
\usepackage[dvips]{graphicx}
\usepackage{verbatim} %
\usepackage{color}
\newcommand{\beq}{\begin{equation}}
\newcommand{\eeq}{\end{equation}}
\newcommand{\bea}{\begin{eqnarray}}
\newcommand{\eea}{\end{eqnarray}}
\usepackage{color}

\newcommand{\bma}{\left(\begin{matrix}}
\newcommand{\ema}{\end{matrix}\right)}
\newcommand{\tom}{\tilde{\omega}}
\newcommand{\om}{\omega}
\def\build#1_#2{\mathrel{\mathop{#1}\limits_{#2}}}
\definecolor{pink}{rgb}{1,0.5,0.5}
\definecolor{violet}{rgb}{1,0,1} 
\definecolor{red}{rgb}{1,0,0}
\definecolor{yellow}{rgb}{0.7,1,0}
\definecolor{orange}{rgb}{1,0.5,0}
\definecolor{white}{rgb}{1,1,1}
\definecolor{blue}{rgb}{0,0,1}
\definecolor{cyan}{rgb}{0,1,1}

\begin{document}

\begin{abstract}
The normal modes of a continuum solid endowed with a random distribution of line defects that behave like elastic strings are described.  These strings interact with elastic waves in the bulk, generating wave dispersion and attenuation. As in amorphous materials, the attenuation as a function of frequency $\omega$ behaves as $\omega^4$ for low frequencies, and, as frequency increases, crosses over to $\omega^2$ and then to linear in $\omega$.  Dispersion is negative in the frequency range where attenuation is quartic and quadratic in frequency. Explicit formulae are provided that relate these properties to the density of string states. { Continuum mechanics can thus be applied both to crystalline materials and their amorphous counterparts at similar length scales.} The possibility of linking this model with the microstructure of amorphous materials is discussed. 
\end{abstract}

\title{Normal modes, and acoustic properties, of an elastic solid with line defects}

\author{Fernando Lund}

\affiliation{\mbox{Departamento de F\'\i sica and CIMAT, Facultad de Ciencias
F\'\i sicas y Matem\'aticas, Universidad de Chile, Santiago, Chile} }


\maketitle

\paragraph{Introduction.} 
The normal modes of a continuum elastic solid can be easily counted using the classical theory of elasticity. However, since there is no intrinsic length scale, an artificial short distance cut-off must be introduced in order to obtain a finite result for the total number of modes of a given material. The Debye model does that, imposing a high frequency cut-off, the Debye frequency $\om_D$, so that the resulting total number of degrees of freedom equals the number of degrees of freedom inferred from the number of atoms in the solid. This provides a firm underpinning, at wavelengths long compared to interatomic spacing, for all properties of solids that depend on the counting on such modes. If the solid is crystalline, a similar counting can also be performed, exploiting the invariance of the system under discrete translations. This counting reduces to that provided by the Debye model at long wavelengths, and provides, as well, a firm underpinning for properties at shorter wavelengths, down to the size of the unit cell.

The situation for amorphous solids, without a discrete translation invariance, has long been unsatisfactory. While at long wavelengths the situation is well described, as expected, by the Debye model, at wavelengths on the order of tens of mean interatomic distances, abundant evidence, from specific heat \cite{sh}, thermal conductivity \cite{thermalc}, Raman scattering \cite{raman}, neutron scattering \cite{neutron}, and inelastic X-ray scattering measurements \cite{ixs}, points to the existence of normal modes with a frequency distribution that is peaked around 0.1-0.2 $\om_D$. This distribution is qualitatively similar for many such materials, and the details, but not the broad features, depend on external parameters such as temperature, density, pressure, as well as chemical and thermal history \cite{bpexp1,bpexp2,bpexp3,bpexp4,bpexp5,bpexp6,bpexp7,bpexp8,bpexp9,bpexp10}. This distribution, dubbed the ``Boson Peak'' (BP), cannot be blamed on a (non-existent) crystalline structure, and deviates without ambiguity from the distribution for a continuum, at frequencies where the continuum approximation works reasonably well in the case of crystals. Much research has been performed in order to provide some rationale for this state of  affairs \cite{teo1,teo2,teo3,teo4,schir2007,teo5,teo6,parshprb2013}. Nevertheless, it does not seem unfair to say that no satisfactory understanding exists yet, although significant insights have been obtained through large scale ($\sim 10^7$ particles) simulations with Lennard-Jones \cite{monacomossa} and soft sphere \cite{schirma13} potentials .

In addition to the above, Ruffl\'e et al. discovered that acoustic attenuation as a function of frequency $\omega$ behaves like $\om^4$ in densified silica\cite{Ruffle2003} and in lithium diborate\cite{Ruffle2006}, and more recent experiments \cite{MonacoGiordano,ruta1,baldiprl,baldijncs,baldiprb,sorbitol} have provided a detailed measurement of the properties of acoustic waves in amorphous materials at wavelengths that probe length scales around the Boson peak. Unexpected behavior for dispersion and attenuation has been uncovered, and the thought naturally comes to mind that whatever the physical mechanism is that gives rise to the Boson peak, it should also explain the acoustic behavior at THz frequencies. This Letter provides one such possible explanation { together with an answer to the question, how can continuum mechanics, at similar length scales, be made to work similarly well for crystals and for their amorphous counterparts.}

\begin{figure}
\includegraphics[width=.8\columnwidth]{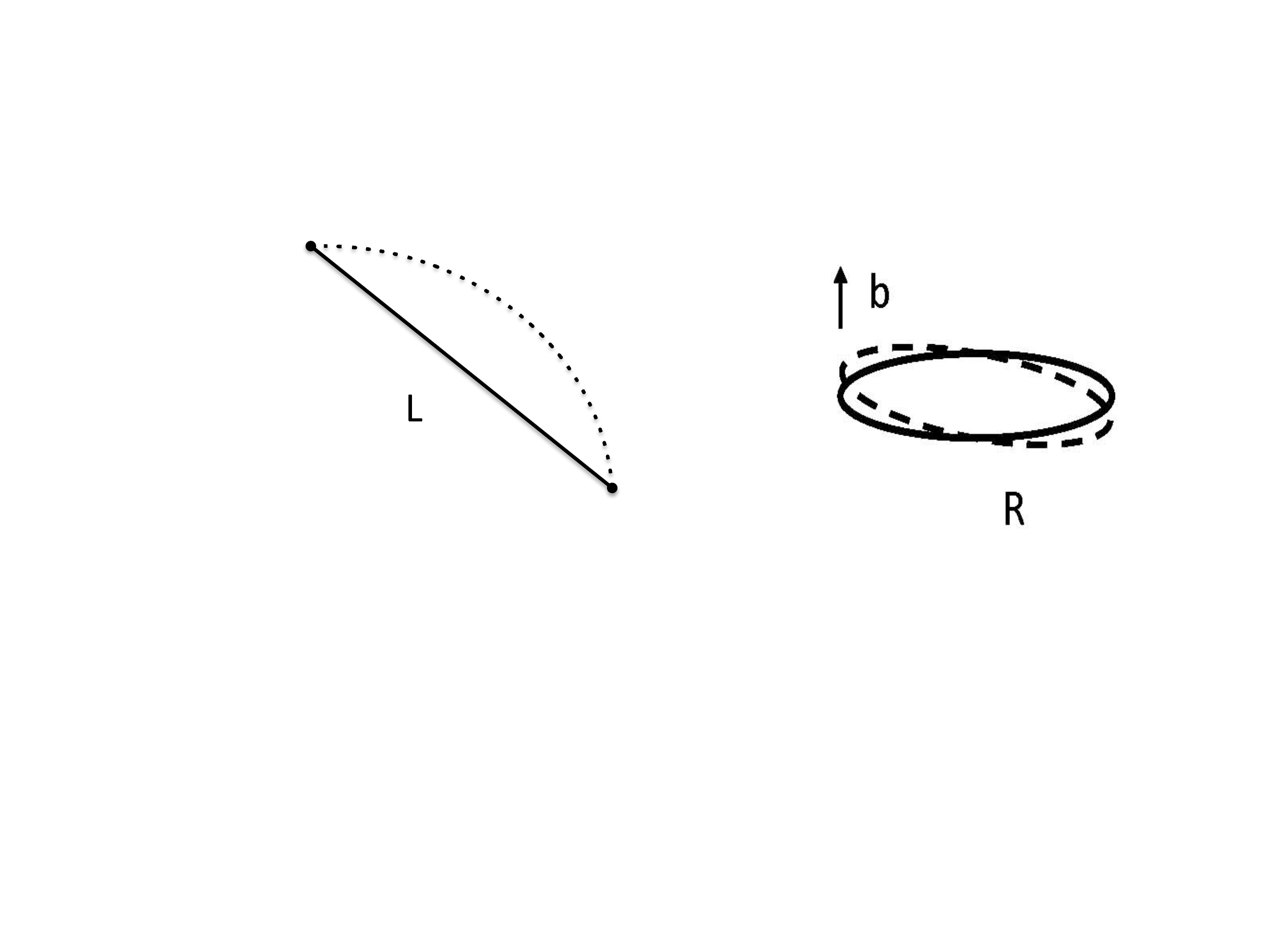}
\caption{Line defects in a continuum elastic solid: Left, an elastic string of length $L$ with pinned ends  that can oscillate around a straight-line equilibrium position. Right, an elastic circular string of radius $R$ that can oscillate around a circular equilibrium position, along the mantle of a circular cylinder with generator $b$. Only the case of a straight line is considered in the present paper; results for the circle are qualitatively similar.}
\label{Figure1}
\end{figure}

\paragraph{The model and its density of states}  At THz frequencies, continuum mechanics provides an adequate description for the vibrational degrees of freedom of crystals. It is proposed that it should also provide an adequate explanation for amorphous materials: Consider then an elastic isotropic solid, as in the Debye model, with $c_L$ and $c_T$ the speed of propagation of longitudinal and transverse waves, respectively. In addition, this solid may have line defects that can be thought of as elastic strings endowed with mass per unit length $m \sim \rho b^2$ and line tension $T \sim \mu b^2$, where $\rho$ is the mass density of the bulk, $\mu$ its shear modulus, and $b$ is a distance on the order of a typical interatomic spacing. These line defects can be pinned segments of length $L$ that can vibrate around a straight equilibrium position, or circular loops that can oscillate, moving along the mantle of a cylinder around a circular equilibrium position (Figure \ref{Figure1}). The precise physical nature of these line defects is delayed for later discussion. The fact that they are defects is used to relate $m$ with $\rho$ and $T$ with $\mu$. As a result, the speed of propagation of elastic waves {\em along} the strings, $c_S$, is of the same order of magnitude as $c_L$ and $c_T$. 

To start with a simple description, assume that only the fundamental mode of vibration of the strings is relevant, ignoring higher harmonics. The strings have a length $L$ with a distribution of such lengths: Let $p(L)dL$ be the number of strings per unit volume with length between $L$ and $L+dL$. The assumption that only the fundamental mode, with frequency $\omega_1$, counts, means there is a one-to-one correspondence between $\omega_1$ and $L$, and leads to a string density of states $g_S(\omega)$ (number of states per unit frequency per unit volume) given by $g_S(\omega) d\omega \equiv p(L) dL$. 
 The total density of states $g(\omega)$ of this model solid will be
\beq
g(\omega) = g_D(\omega) + g_S(\omega)
\eeq
where $g_D(\omega) = 3\omega^2/2c^3\pi^2$ is the Debye distribution, with $3c^{-3}\equiv c_L^{-3} +2c_T^{-3}$. For later convenience, we shall normalize the distribution $g_S$ to $g_D$, and frequencies to the Debye frequency:
\beq
g_S(\om) \equiv \tilde g (\tilde{\om}) g_D(\omega)
\label{reduced}
\eeq
where $\tilde{\om} \equiv \om/\om_D$  

\paragraph{Acoustic properties} 
An elastic string will oscillate in response to loading by an elastic wave. These oscillations will, in turn, generate secondary, scattered waves, a process that has been studied in great detail by Maurel et al. \cite{M4,maureljap,Maurel2009D}. The dynamics is described by the string displacements $X(s,t)$ away from its equilibrium position, where $s$ is a Lagragean parameter to label string points, and $t$ is time . The displacement $X$  satisfies  an elastic string equation \cite{lund88}
\begin{equation}
m \ddot{X}( s ,t) +B\dot{X}( s ,t)- T X''( s ,t)= \mu b \; {\mathsf
  M}_{lk}bla_l u_k( \vec X ,t)  , 
\label{eqmouv2}
\end{equation}
with  $B$ a phenomenological viscous damping coefficient. An overdot denotes differentiation with respect to time $t$, and a prime, with respect to $s$. The motion is along a plane spanned by the tangent $\hat{\tau}$ to the string, and a normal $\hat t$. Until further notice, $B$ will be an adjustable parameter of the model. The binormal is $\hat n \equiv \hat{\tau} \times \hat t$ and ${\mathsf M}_{lk}\equiv t_l n_k+t_kn_l.$ The right-hand-side of (\ref{eqmouv2}) is the Peach-Koehler force for line dislocations in elastic continua. This coupling ensures that only the shear modulus, and not the bulk modulus, will become frequency dependent, a fact that has been observed in the numerical simulations of Maruzzo et al. \cite{schirma13}.

\begin{figure}[h!]
\includegraphics[width=.8\columnwidth]{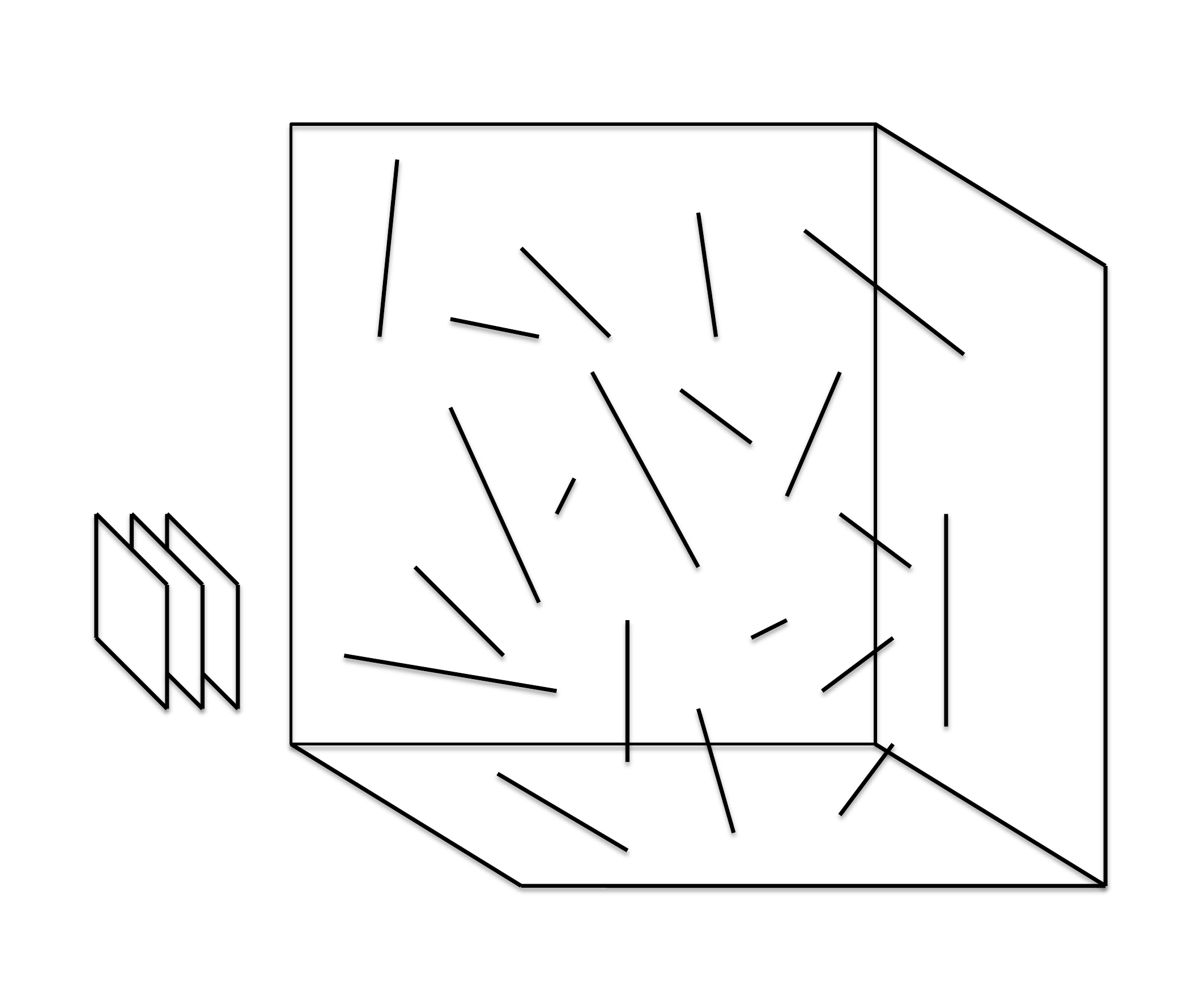}
\caption{Line defects, randomly distributed and oriented, have an effect on the properties of an acoustic wave, here depicted as a set of planes incident from the left. Attenuation is given by Eqn.(\ref{attvsbp}), and velocity dispersion by Eqn. (\ref{dispersion}). Their lengths have a distribution that, through the frequency of their fundamental mode of oscillation, translates to a frequency distribution.}
\label{Figure2}
\end{figure}

We now wish to consider the interaction of acoustic waves with many such strings, randomly distributed, and randomly oriented (Figure \ref{Figure2}). We assume all positions equally likely, as well as all orientations. Lengths, however, have a distribution $p(L)$. An elastic wave progressing through this medium will attenuate,  and will propagate with  a frequency-dependent velocity, both for longitudinal as well as transverse polarizations. In the following we shall treat the transverse case, for which the effect is stronger \cite {M4}. This is also in agreement with the measurements of Chumakov et al. \cite{chumaBP} that link the boson peak of a glass with the transverse acoustic van Hove singularity of its crystalline counterpart. The attenuation coefficient $\Gamma_T$ is given by (in units of frequency)
\beq
\Gamma_T =  c_T \int \sigma_T (L) p(L) dL
\label{singlescat}
\eeq
where $\sigma$ is the total scattering cross section for a transverse wave by a single defect. It will, in general, depend on the relative orientation of the string and the incident wave, so we shall take the average value. The resulting average cross section is \cite{M4}, 
\beq
\sigma_T(L)  =    \frac{2 L^2}{25 \pi^5} \left( \frac{\rho b^2}{m} \right)^2 \frac{\omega^4}{(\omega^2 - (\pi c_s/L)^2)^2 + \omega^2 B^2 /m^2}
\label{cross}
\eeq
where  $c_S^2 = T/m$.  Using (\ref{reduced}), (\ref{cross}), $\omega_1^2 = (\pi c_s/L)^2 - (B/2m)^2$,  and $m = \rho b^2 / \pi$,  Eqn. (\ref{singlescat}) becomes 
\beq
{\tilde{\Gamma}}_T (\omega) = \frac{3 {\tilde{\omega}}^4}{ 25 \pi^3}  \int \frac{\tilde g (\tilde{\omega}_1) {\tilde{\om}}_1^2 d\tilde{\omega}_1}{{\cal A}({\tilde{\om}}_1 ) {\cal B}(\tom ,{\tilde{\om}}_1 )} \, ,
\label{attvsbp}
\eeq
with $\tilde b \equiv B/m\omega_D, {\tilde{\Gamma}}_T \equiv \Gamma_T /\om_D$, and 
\bea
{\cal A}({\tilde{\om}}_1 ) & \equiv &  {\tilde{\om_1}}^2 + {\tilde b}^2 /4 \\
{\cal B}(\tom ,{\tilde{\om}}_1 ) & \equiv  & ({\tilde{\om}}^2 - {\tilde{\om_1}}^2 - {\tilde b}^2 /4 )^2 + {\tilde{\om}}^2 {\tilde b}^2
\eea

Equation (\ref{attvsbp})  establishes a clear relation between the degrees of freedom responsible for the Boson peak and the degrees of freedom responsible for acoustic attenuation: Indeed, they are one and the same. At low frequencies, there is a $\tilde{\omega}^4$ Rayleigh behavior, as expected for attenuation produced by the scattering from objects much smaller than wavelength, and as observed experimentally \cite{MonacoGiordano,baldijncs}. Also, the magnitude of the attenuation is proportional to the magnitude of the BP and its exact functional behavior depends on the parameters of the density of states, plus the viscous damping $\tilde b$.

The scattering  by strings also modifies the coherent behavior of acoustic waves, a phenomenon that has been studied by Maurel et al. \cite{M5}. Their result, obtained through a multiple scattering formalism, is that the velocity of propagation $c_T$ is renormalized to an effective velocity $v_T$ given by 
\beq
v_T   \approx   c_T \left( 1 + \frac{1}{\pi^2}  \int \frac{(\tom^2 -{\cal A}(\tom_1)) \, \tilde g (\tilde{\omega}_1) \, \tom_1^2 \, d\tom_1}{{\cal A}(\tom_1) ^{1/2} {\cal B} (\tom , \tom_1) }  \right) 
\label{dispersion}
\eeq
 in terms of dimensionless, scaled, variables. The string degrees of freedom introduce a dispersion term into the effective velocity of acoustic waves. Eqn. (\ref{dispersion}) establishes a clear relation between the dispersion term and the density of states responsible for the Boson peak. Note that the dispersion changes sign at about the same frequency that the attenuation (\ref{attvsbp}) has a maximum, and the amount of dispersion is proportional to the amplitude of the BP. 

\paragraph{Possible values for model parameters}  
In order to get a better understanding of the relation between line defects, density of states and acoustics that is being proposed, consider the following density of states.
\beq
\tilde g (\tilde{\omega})  =  \alpha \left[ 1-\delta^{-2} \left( \tom -\beta \right)^2 \right] 
\label{ges}
\eeq
when $(\omega/ \omega_D -\beta)^2 < \delta^2$, and zero otherwise. This is an inverted parabola, where $\alpha$, $\beta$ and $\delta$ are dimensionless, adjustable, parameters characterizing the amplitude, position and width, respectively,  of the BP.

\begin{figure}
\includegraphics[width=.8\columnwidth]{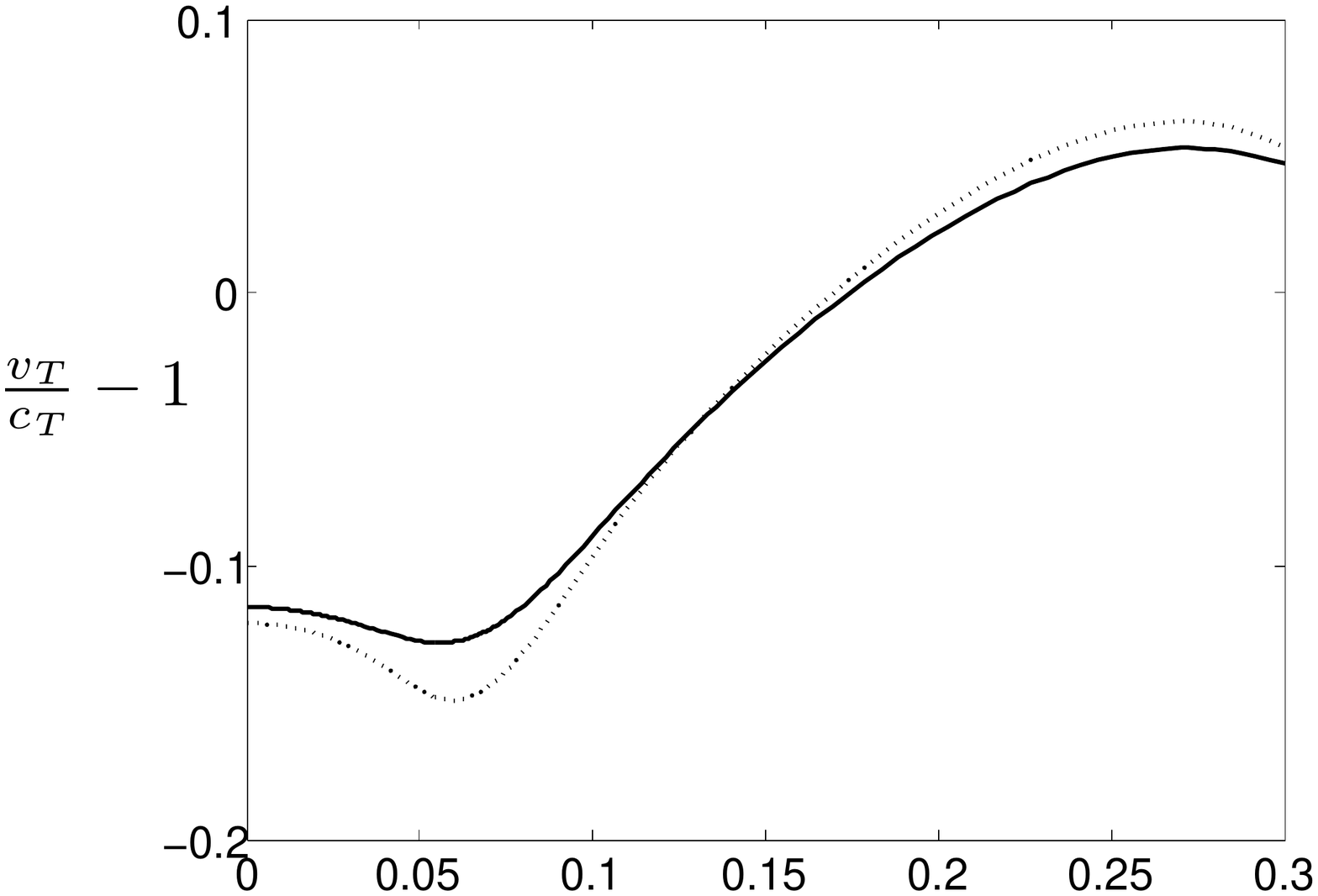}
\includegraphics[width=.8\columnwidth]{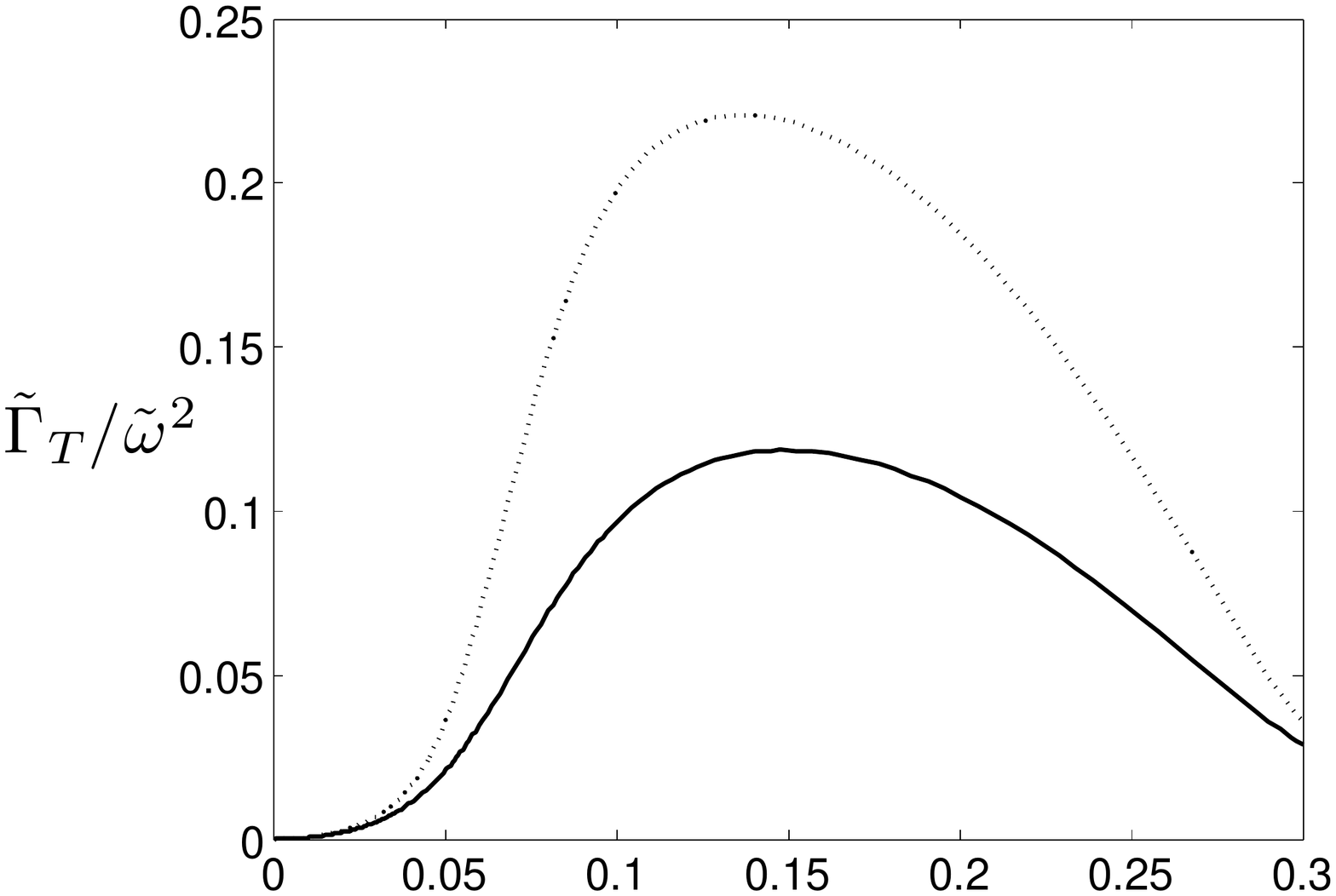}
\includegraphics[width=.8\columnwidth]{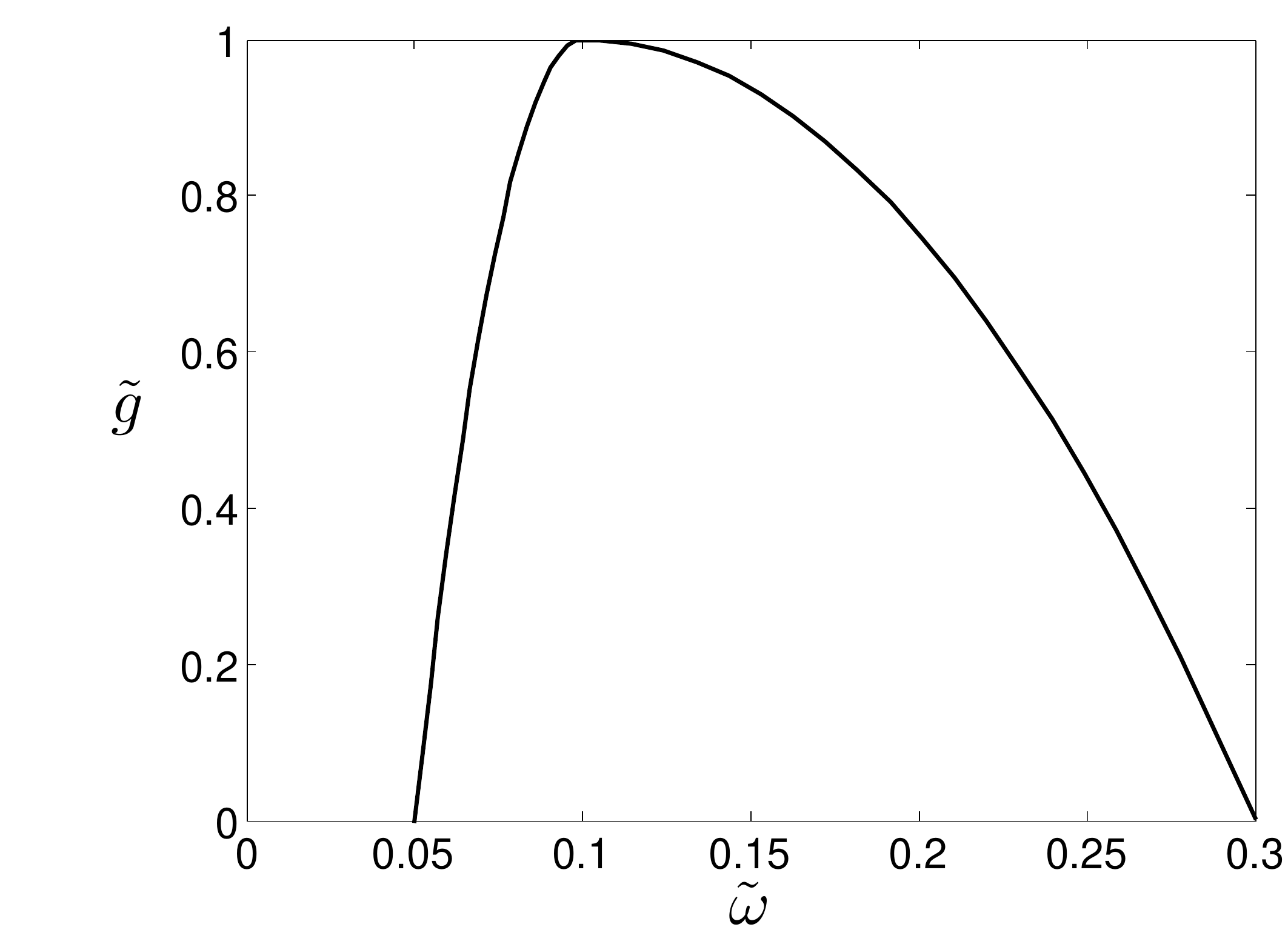}
\caption{Bottom panel: Density of states associated with a distribution of strings of varying lengths, given by (\ref{ges}) with $\alpha = 1$, $\beta = 0.1$ and $\delta = 0.05-0.2$. Middle panel: Acoustic attenuation generated by these strings, given by Eqn. (\ref{attvsbp}), with $\tilde b =0.05$ (solid line) and $\tilde b =0.03$ (dotted line) . Upper panel: velocity dispersion generated by the same distribution of strings, given by Eqn. (\ref{dispersion}). All frequencies are in units of $\om_D$.}
\label{Figure3}
\end{figure}

Typically, the magnitude of the BP is a factor of about 2 over the Debye DOS, it is located at about 0.1 $\omega_D$, with a width of about 0.05 $\omega_D$. So, in (\ref{ges}) we let $\alpha = 1, \beta =0.1$, and $\delta \sim 0.05-0.2$. Next an estimate must be given for $\tilde b$. In principle, it could be different for different string lengths, i.e., a function of $\tom$. To start with the simplest possibility, assume that it is the same for all strings. As discussed more fully in the next section, we wish the strings to be a description of the collective, coherent, behavior of several tens of atoms, linearly arranged. Consequently, we take $\tilde b \sim 0.05-0.03$. Figure \ref{Figure3} shows a density of states given by (\ref{ges}), the resulting acoustic attenuation, given by  (\ref{attvsbp}), and velocity dispersion, given by (\ref{dispersion}), for these values of the parameters. First, the magnitude of the effects appear to be in order-of-magnitude agreement with experimental observations: Attenuation at the level of 1\% of $\om_D$ and a velocity dip, due to negative dispersion, at the level of 10\% of the low frequency value.  Second, the attenuation behaves as $\tom^4$ at low frequencies, then crosses over to a $\tom^2$ behavior near the BP, and then goes over to a linear $\tom$ behavior beyond. This is also in agreement with observations. The behavior of the acoustic velocity as a function of frequency starts at a constant value, then has a negative dispersion (the velocity decreases as the frequency increases) as the BP frequencies are approached, and then acoustic velocity increases with frequency. 

How many line defects are present, given the parameter values assumed so far, and what is their mean length? The maximum of the distribution (\ref{ges}) is at 0.1 $\om_D$,  so string lengths are distributed roughly between 5 and 20 interatomic spacings with a peak at 10.  
The density $n$ of line defects (number per unit volume) is
\[
n= \int p(L) dL = \int g_S(\omega) d\omega \approx 10^{-3} \frac{3}{2\pi^2} \left( \frac{\omega_D}{c} \right)^3
\]
which means there is about one string per cube of side 10 interatomic spacings.

\paragraph{Possibility of a relation to the microstructure of  amorphous materials} 
The analysis that has been carried out here is wholly within a continuum mechanics approximation. How could the results be linked with the structure of amorphous materials at the atomic scale? In the case of crystals, Volterra dislocations in a continuum seamlessly blend to dislocations in a crystalline structure, the latter providing a specific value for their Burgers vector. These line defects, at least in small enough numbers, break the short range order but not the long range order. The existence of crystallographic slip planes allows for the large scale motion of these defects that explains crystal plasticity. Could line defects, microscopically well defined, exist in amorphous materials, without slip planes? Point defects are well known and are responsible for important technological properties \cite{ovsh}. Indeed, the possibility of specific vitreous state defects as a source of BP vibrations has been explored \cite{angelld1,angelld2,angelld3}. Could such point defects be arranged in a necklace to give rise to a line defect? At least, there does not appear to be an argument of a general nature (say, energetic, entropic or topologic) to rule out their existence.{  Alternatively, line defects arranged in knots would also destroy the slip planes \cite{referee}.} Karpov\cite{karpov} has considered the possibiility that acoustic waves propagate along closed loop like trajectories, as a consequence of the randomness of the sound velocity, leading to a density of states that competes with that of phonons, and Novikov and Surotsev\cite{novisuro} have shown that Raman scattering in glasses is consistent with boson peak vibrations belonging to a one dimensional spatial structure. 

From a different point of view, it has been well established, through numerical simulation, that atomic displacements in amorphous Lennard Jones, Silica and Silicon are split into affine and non-affine modes \cite{tanprb2002,tanprb2004,tanprb2005,tsampre2009,leonfprl2006,Tanepl2009}. The possibility arises that the line defects envisaged here are responsible for the non-affine modes. In addition, Vural \& Legget \cite{VL} have developed a formalism that blames the low temperature properties of amorphous materials on a splitting of acoustic properties into phonon and non-phonon modes, a distinction that is similar in spirit to what is carried out here. Finally, signatures of collective string-like motion in supercooled liquids in computer simulations have also been found\cite{kob,starr}. A microscopic link to the line defects envisaged in this paper could provide a useful tool to understand the behavior of the BP as a function of external control parameters such as temperature and pressure.

\paragraph{Discussion.}
A model has been proposed, within the framework of continuum mechanics, that links the BP in amorphous materials to the properties of acoustic waves { in interaction with line defects. This approach bears some similarity to the time-honored treatment of the quantum hydrodynamics of superfluid helium and atomic  Bose-Einstein condensates through (quantized) vortex lines \cite{vor1,vor2,vor3}.} { It also provides an answer to the long-standing question, how to apply continuum mechanics, at similar length scales, both to crystals and their amorphous counterparts.} { In the present work,} the degrees of freedom responsible for the BP are taken to be the vibrations of line defects around an equilibrium position; they are also responsible for acoustic attenuation and dispersion. Figure \ref{Figure3} shows an example of the effect of a given DOS, for two values of the parameters, on attenuation and dispersion. As expected, a larger number of defects provides a higher effect. Dispersion behaves as $\om^4$ at low frequencies, then crosses over to $\om^2$ at frequencies $\om_{\rm BP}$ comparable to the BP, and then, at even higher frequencies, to linear in $\om$. The velocity of acoustic waves has a negative dispersion at low frequencies and then increases as a function of frequency around $\om_{\rm BP}$. Notice that the defects have an effect even at low frequencies, suggesting the model can be tested, say, using Resonant Ultrasound Sectroscopy (RUS), a tool that has been used to measure dislocation densities in polycrystalline materials \cite{Barra2009,rus}, { or hyper sound damping in the subteraherz range, as measured in vitreous silica \cite{m1,m2}}.   Additional possible topics for further research include a numerical, atomistic, study of the hypothesized linear defects; the effect they would have on thermal and electrical conductivity properties, particularly at very low temperatures where quantum effects would be dominant; and the role they may or may not have in plasticity properties and in the glass transition.

\paragraph{Acknowledgements.}
I wish to thank A. Tanguy for enlightening discussions. The support of Fondecyt Grant 1130382 and ANR-Conicyt Grant PROCOMEDIA is gratefully acknowledged.

\end{document}